\documentclass[12pt,twoside,dvips]{article}
\usepackage{amssymb}
\usepackage{amsfonts}
\usepackage{amsmath}
\usepackage[spanish,english]{babel}
\usepackage{graphicx}

\begin{document}

\title{The Landau Pole and $Z^{\prime }$ decays in the $331$ bilepton model}
\author{R. Mart\'{\i}nez$\thanks{%
e-mail: remartinezm@unal.edu.co}$ \ and F. Ochoa$\thanks{%
e-mail: faochoap@unal.edu.co}$ \and Departamento de F\'{\i}sica, Universidad
Nacional, \\
Bogot\'{a}-Colombia}
\maketitle

\begin{abstract}
We calculate the decay widths and branching ratios of the extra neutral
boson $Z^{\prime}$ predicted by the 331 bilepton model in the framework of
two different particle contents. These calculations are performed taken into
account oblique radiative corrections, and Flavor Changing Neutral Currents
(FCNC) under the ansatz of Matsuda as a texture for the quark mass matrices.
Contributions of the order of $10^{-1}-10^{-2}$ are obtained in the
branching ratios, and partial widths about one order of magnitude bigger in
relation with other non- and bilepton models are also obtained. A
Landau-like pole arise at $3.5$ TeV considering the full particle content of
the minimal model (MM), where the exotic sector is considered as a
degenerated spectrum at $3$ TeV scale. The Landau pole problem can be avoid
at the TeV scales if a new leptonic content running below the threshold at $%
3 $ TeV is implemented as suggested by other authors.
\end{abstract}

\vspace{-0.3cm}

\vspace{-5mm}

\section{Introduction}

The models with gauge symmetry $SU(3)_{c}\otimes SU(3)_{L}\otimes U(1)_{X},$
also called 331 models, arise by enlarging the symmetry group where the
Standar Model (SM) is embedded as an effective theory at low energy. Based
on the criterion of cancellation of chiral anomalies \cite{anomalias}, these
models have generated new expectations and possibilities of new physics at
the TeV scale whose predictions are sensitive to experimental observation
and will be of great interest in the next generation of colliders (LHC, ILC) 
\cite{godfrey} at the TeV energy scales. Among the most remarkable features
of these models, we have that they arise as an interesting alternative to
explain the origin of generations \cite{Frampton2}, they leads to the
prediction of new neutral and charged vector bosons with observable
consequences at low and high energies, and they can predict the charge
quantization for a three family model even when neutrino masses are added 
\cite{Pires}.

Although cancellation of anomalies leads to some required conditions \cite%
{fourteen}, such criterion alone still permits an infinite number of 331
models. In these models, the electric charge is defined in general as a
linear combination of the diagonal generators of the group

\begin{equation}
Q=T_{3}+\beta T_{8}+XI,  \label{charge}
\end{equation}%
where the value of the $\beta $ parameter determines the fermion assignment
and, more specifically, the electric charges of the exotic spectrum. Hence,
it is customary to use this quantum number to classify the different 331
models. There are two main versions of 331 models corresponding to $\beta
=-1/\sqrt{3}$ \cite{fourteen, twelve} which leads to non-exotic charges, and 
$\beta =-\sqrt{3}$ \cite{ten} which accounts for doubly charged bilepton
bosons. An extensive and detailed study of models with $\beta $ arbitrary
have been carried out in ref. \cite{331us} for the scalar sector and in ref. 
\cite{beta-arbitrary} for the fermionic and gauge sector.

The group structure of these models leads, along with the SM neutral boson $%
Z,$ to the prediction of an additional current associated to the extra
neutral boson $Z^{\prime }.$ It is possible to obtain constraints through
direct production of the $Z^{\prime }$ boson (for center-of-mass energy $%
\sqrt{s}\approx M_{Z^{\prime }}>M_{Z}$), where the study of decay widths
provide information about possible $Z^{\prime }-$detection in future
experimental measurements. These decay widths are controlled by the $%
SU(3)_{L}\otimes U(1)_{X}$ coupling constants denoted by $g_{L}$ and $g_{X}$%
, respectively, and which follows the relation

\begin{equation}
\frac{g_{X}^{2}}{g_{L}^{2}}=\frac{S_{W}^{2}}{1-(1+\beta ^{2})S_{W}^{2}},
\label{POLE}
\end{equation}%
where $S_{W}=\sin \theta _{W}$ with $\theta _{W}$ the Weinberg angle. The
Eq. \ref{POLE} exhibit a Landau pole when $S_{W}^{2}(\mu )=1/(1+\beta ^{2}),$
where the coupling constant $g_{X}$ becomes infinite, and the models lose
their perturbative character at some energy scale $\mu .$ In particular, the
model with $\beta =-\sqrt{3}$ shows a Landau pole when $S_{W}^{2}=1/4$ at
the scale $\mu \approx 4$ TeV, which is near to the bounds associated to the 
$Z^{\prime }-$mass predicted by this model \cite{family-dependence}. Thus,
this version of the 331 models points to a nonperturbative regime at the $%
Z^{\prime }$ peak. This problem has already been considered in ref. \cite%
{pole1} with and without supersymmetry of the bilepton model. In particular,
a possible solution of this puzzle was proposed in ref. \cite{pole2} by
introducing an additional particle content, which change the behaviour of
the running Weinberg angle and restores the perturbative feature of the
model. The new particle content is composed by three exotic lepton triplets,
one exotic scalar triplet and two exotic scalar doublets. In this extended
model (EM), the usual exotic particles predicted by the minimal 331 model
(MM) remains as heavy particles which are decoupled below the scale $\mu
_{331}$ where the symmetry breaking $SU(3)_{L}\times U(1)_{X}\rightarrow
SU(2)_{L}\times U(1)_{Y}$ takes place$,$ so that they are not considered as
active degrees of freedom below the $Z^{\prime }$ energy scale. However,
some phenomenological studies estimates that the exotic quarks have a mass
in the range $1.5-4$ TeV \cite{Roberto-Sampayo}, and they could enter as
active degree of freedom at the $Z^{\prime }$ scale.

In this work we study the behaviour of the running Weinberg angle for
different particle content in the 331 bilepton model, where the exotic
particles define a new threshold energy scale below $M_{Z^{\prime }}$.
Later, we obtain the partial widths and the branching decays of the $%
Z^{\prime }$ boson in the 331 bilepton model for two perturbative particle
content, where the running coupling constants at $M_{Z^{\prime }}$ scale is
taken into account in agreement with the new particle content. This paper is
organized as follows. Subsection 2.1 is devoted to summarize the minimal 331
bilepton model (MM) and the behaviour of the running Weinberg angle. In
Subsec. 2.2, we indicate the most remarkable features of the extended 331
bilepton model (EM) in order to obtain a decreasing running Weinberg angle.
In Subsec. 2.3, we consider a modified extended 331 bilepton model (MEM),
where we include the exotic spectrum of the minimal model as active degree
of freedom below the $Z^{\prime }$ scale. Finally, in Sec. 3 we calculate de
partial decays and branching ratios of the $Z^{\prime }$ boson in the
framework of the EM and MEM bilepton models, where effects of Flavor
Changing Neutral Currents (FCNC) and oblique radiative corrections are taken
into account.

\section{Perturbative and nonperturbative bilepton models}

In this section we show different particle contents in the 331 bilepton
model, which control the behavior of the Weinberg angle and the coupling
constants with the increasing of the energy.

\subsection{The minimal 331 bilepton model (MM)}

The minimal 331 fermionic structure for three families is shown in table \ref%
{tab:espectro} for the bilepton model corresponding to $\beta =-\sqrt{3}$,
where all leptons transform as $(\mathbf{1,3,X}_{\ell }^{L})$ and $(\mathbf{%
1,1,X}_{\ell ^{\prime }}^{R})$ under $\left(
SU(3)_{c},SU(3)_{L},U(1)_{X}\right) ,$ with $\mathbf{X}_{\ell }^{L}$ and $%
\mathbf{X}_{\ell ^{\prime }}^{R}$ the $U(1)_{X}$ values associated to the
left- and right-handed leptons, respectively; while the quarks transform as $%
(\mathbf{3,3}^{\ast }\mathbf{,X}_{q_{m^{\ast }}}^{L})$, $(\mathbf{3}^{\ast }%
\mathbf{,1,X}_{q_{m^{\ast }}^{\prime }}^{R})$ for the first two families,
and $(\mathbf{3,3,X}_{q_{3}}^{L})$, $(\mathbf{3}^{\ast }\mathbf{,1,X}%
_{q_{3}^{\prime }}^{R})$ for the third family, where $\mathbf{X}_{q_{m^{\ast
}}}^{L},\mathbf{X}_{q_{3}}^{L}$ and $\mathbf{X}_{q_{m^{\ast }}^{\prime
}}^{R},\mathbf{X}_{q_{3}^{\prime }}^{R}$ correspond to the $U(1)_{X}$ values
for left- and right-handed quarks. We denote $\mathbf{X}_{q_{3}}^{L}$ and $%
\mathbf{X}_{q_{m^{\ast }}}^{L}$ as the values associated to the $SU(3)_{L}$
space under representation $\mathbf{3}$ and $\mathbf{3}^{\ast }$,
respectively. The quantum numbers $\mathbf{X}_{\psi }$ for each
representation are given in the third column from table \ref{tab:espectro},
where the definition of the electric charge in Eq. \ref{charge} has been
used, demanding charges of $2/3$ and $-1/3$ to the up- and down-type quarks,
respectively, and charges of -1,0 for the charged and neutral leptons. We
recognize three different possibilities to assign the physical quarks in
each family representation as shown in table \ref{tab:combination} \cite%
{Mohapatra}. At low energy, the three models from table \ref{tab:combination}
are equivalent and there are not any phenomenological feature that allow us
to detect differences between them. In fact, they must reduce to the SM
which is an universal family model in $SU(2)_{L}.$ However, through the
couplings of the three families to the additional neutral current ($%
Z^{\prime }$) and the introduction of a mixing angle between $Z$ and $%
Z^{\prime },$ it is possible to recognize differences among the three models
at the electroweak scale \cite{family-dependence,z2-decay}.

\begin{table}[tbp]
\begin{center}
\begin{equation*}
\begin{tabular}{c|c|c}
\hline
$representation$ & $Q_{\psi }$ & $X_{\psi }$ \\ \hline\hline
$\ 
\begin{tabular}{c}
$q_{m^{\ast }L}=\left( 
\begin{array}{c}
d_{m^{\ast }} \\ 
-u_{m^{\ast }} \\ 
J_{m^{\ast }}%
\end{array}%
\right) _{L}:\mathbf{3}^{\ast }$ \\ 
\\ 
\\ 
$d_{m^{\ast }R};$ $u_{m^{\ast }R};$ $J_{m^{\ast }R}:\mathbf{1}$%
\end{tabular}%
\ $ & 
\begin{tabular}{c}
$\left( 
\begin{array}{c}
-1/3 \\ 
2/3 \\ 
-4/3%
\end{array}%
\right) $ \\ 
\\ 
$-1/3;$ $2/3;$ $-4/3$%
\end{tabular}
& 
\begin{tabular}{c}
\\ 
$-1/3$ \\ 
\\ 
\\ 
$-1/3,2/3,-4/3$%
\end{tabular}
\\ \hline\hline
\begin{tabular}{c}
$q_{3L}=\left( 
\begin{array}{c}
u_{3} \\ 
d_{3} \\ 
J_{3}%
\end{array}%
\right) _{L}:\mathbf{3}$ \\ 
\\ 
$u_{3R};$ $d_{3R};$ $J_{3R}:\mathbf{1}$%
\end{tabular}
& 
\begin{tabular}{c}
$\left( 
\begin{array}{c}
2/3 \\ 
-1/3 \\ 
5/3%
\end{array}%
\right) $ \\ 
\\ 
$2/3;$ $-1/3;$ $5/3$%
\end{tabular}
& 
\begin{tabular}{c}
\\ 
$2/3$ \\ 
\\ 
\\ 
$2/3,-1/3,5/3$%
\end{tabular}
\\ \hline\hline
\begin{tabular}{c}
$\ell _{jL}=\left( 
\begin{array}{c}
\nu _{j} \\ 
e_{j} \\ 
E_{j}^{+}%
\end{array}%
\right) _{L}:\mathbf{3}$ \\ 
\\ 
$e_{jR};$ $E_{jR}^{-Q_{1}}$%
\end{tabular}
& 
\begin{tabular}{c}
$\left( 
\begin{array}{c}
0 \\ 
-1 \\ 
1%
\end{array}%
\right) $ \\ 
\\ 
$-1;$ $1$%
\end{tabular}
& 
\begin{tabular}{c}
\\ 
$0$ \\ 
\\ 
\\ 
$-1,$ $1$%
\end{tabular}
\\ \hline
\end{tabular}%
\end{equation*}%
\end{center}
\caption{\textit{Fermionic content for three generations with\ }$\protect%
\beta =-\protect\sqrt{3}$\textit{. We take} $m^{\ast }=1,2$, \textit{and} $%
j=1,2,3$}
\label{tab:espectro}
\end{table}

For the scalar sector described by Table \ref{tab:quince}, we introduce the
triplet field $\chi $ with vacuum expectation value (VEV) $\left\langle \chi
\right\rangle ^{T}=\left( 0,0,\nu _{\chi }\right) $, which induces the
masses to the third fermionic components. In the second transition it is
necessary to introduce two triplets$\;\rho $ and $\eta $ with VEV $%
\left\langle \rho \right\rangle ^{T}=\left( 0,\nu _{\rho },0\right) $ and $%
\left\langle \eta \right\rangle ^{T}=\left( \nu _{\eta },0,0\right) $ in
order to give masses to the quarks of type up and down, respectively.

\begin{table}[tbp]
\begin{equation*}
\begin{tabular}{c|c|c}
\hline
Representation $A$ & Representation $B$ & Representation $C$ \\ \hline\hline
$%
\begin{tabular}{c}
$q_{mL}=\left( 
\begin{array}{c}
d,s \\ 
-u,-c \\ 
J_{1},J_{2}%
\end{array}%
\right) _{L}:\mathbf{3}^{\ast }$ \\ 
$q_{3L}=\left( 
\begin{array}{c}
t \\ 
b \\ 
J_{3}%
\end{array}%
\right) _{L}:\mathbf{3}$%
\end{tabular}%
\ $ & $%
\begin{tabular}{c}
$q_{mL}=\left( 
\begin{array}{c}
d,b \\ 
-u,-t \\ 
J_{1},J_{3}%
\end{array}%
\right) _{L}:\mathbf{3}^{\ast }$ \\ 
$q_{3L}=\left( 
\begin{array}{c}
c \\ 
s \\ 
J_{2}%
\end{array}%
\right) _{L}:\mathbf{3}$%
\end{tabular}%
\ $ & $%
\begin{tabular}{c}
$q_{mL}=\left( 
\begin{array}{c}
s,b \\ 
-c,-t \\ 
J_{2},J_{3}%
\end{array}%
\right) _{L}:\mathbf{3}^{\ast }$ \\ 
$q_{3L}=\left( 
\begin{array}{c}
u \\ 
d \\ 
J_{1}%
\end{array}%
\right) _{L}:\mathbf{3}$%
\end{tabular}%
\ $ \\ \hline
\end{tabular}%
\end{equation*}%
\caption{\textit{Three different assignments for the $SU(3)_{L}$ family
representation of quarks}}
\label{tab:combination}
\end{table}

\begin{table}[tbp]
\begin{center}
$%
\begin{tabular}{c|c|c}
\hline
& $Q_{\Phi }$ & $X_{\Phi }$ \\ \hline\hline
$\chi =\left( 
\begin{array}{c}
\chi _{1}^{-} \\ 
\chi _{2}^{-\text{ }-} \\ 
\xi _{\chi }+\nu _{\chi }\pm i\zeta _{\chi }%
\end{array}%
\right) $ & $\left( 
\begin{array}{c}
-1 \\ 
-2 \\ 
0%
\end{array}%
\right) $ & $-1$ \\ \hline
$\rho =\left( 
\begin{array}{c}
\rho _{1}^{+} \\ 
\xi _{\rho }+\nu _{\rho }\pm i\zeta _{\rho } \\ 
\rho _{3}^{++}%
\end{array}%
\right) $ & $\left( 
\begin{array}{c}
1 \\ 
0 \\ 
2%
\end{array}%
\right) $ & $1$ \\ \hline
$\eta =\left( 
\begin{array}{c}
\xi _{\eta }+\nu _{\eta }\pm i\zeta _{\eta } \\ 
\eta _{2}^{-} \\ 
\eta _{3}^{+}%
\end{array}%
\right) $ & $\left( 
\begin{array}{c}
0 \\ 
-1 \\ 
1%
\end{array}%
\right) $ & $0$ \\ \hline
\end{tabular}%
\ \ $%
\end{center}
\caption{\textit{Scalar spectrum that break the 331 symmetry and give masses
to the fermions.}}
\label{tab:quince}
\end{table}

In the gauge boson spectrum associated with the group $SU(3)_{L}\otimes
U(1)_{X},$ we have the charged sector

\begin{equation}
W_{\mu }^{\pm }=\frac{1}{\sqrt{2}}\left( W_{\mu }^{1}\mp iW_{\mu
}^{2}\right) \ ;\ K_{\mu }^{\pm 1}=\frac{1}{\sqrt{2}}\left( W_{\mu }^{4}\mp
iW_{\mu }^{5}\right) \ ;\ K_{\mu }^{\pm 2}=\frac{1}{\sqrt{2}}\left( W_{\mu
}^{6}\mp iW_{\mu }^{7}\right)
\end{equation}

and the neutral sector that corresponds to the photon, the $Z$ and the $%
Z^{\prime }$ bosons

\begin{eqnarray}
A_{\mu } &=&S_{W}W_{\mu }^{3}+C_{W}\left( -\sqrt{3}T_{W}W_{\mu }^{8}+\sqrt{%
1-3T_{W}^{2}}B_{\mu }\right) ,  \notag \\
Z_{\mu } &=&C_{W}W_{\mu }^{3}-S_{W}\left( -\sqrt{3}T_{W}W_{\mu }^{8}+\sqrt{%
1-3T_{W}^{2}}B_{\mu }\right) ,  \notag \\
Z_{\mu }^{\prime } &=&-\sqrt{1-3T_{W}^{2}}W_{\mu }^{8}-\sqrt{3}T_{W}B_{\mu },
\end{eqnarray}%
where the Weinberg angle is defined as 
\begin{equation}
S_{W}=\sin \theta _{W}=\frac{g_{X}}{\sqrt{g_{L}^{2}+4g_{X}^{2}}},\quad
T_{W}=\tan \theta _{W}=\frac{g_{X}}{\sqrt{g_{L}^{2}+3g_{X}^{2}}}
\end{equation}%
Further, a small mixing angle between the two neutral currents $Z_{\mu }$
and $Z_{\mu }^{\prime }$ appears with the following mass eigenstates \cite%
{beta-arbitrary}

\begin{eqnarray}
Z_{1\mu } &=&Z_{\mu }C_{\theta }+Z_{\mu }^{\prime }S_{\theta };\quad Z_{2\mu
}=-Z_{\mu }S_{\theta }+Z_{\mu }^{\prime }C_{\theta };  \notag \\
\tan \theta &=&\frac{1}{\Lambda +\sqrt{\Lambda ^{2}+1}};\quad \Lambda =\frac{%
-2S_{W}C_{W}^{2}g_{X}^{2}\nu _{\chi }^{2}+\frac{3}{2}S_{W}T_{W}^{2}g_{L}^{2}%
\left( \nu _{\eta }^{2}+\nu _{\rho }^{2}\right) }{g_{L}g_{X}T_{W}^{2}\left[
-3\sqrt{3}S_{W}^{2}\left( \nu _{\eta }^{2}+\nu _{\rho }^{2}\right)
+C_{W}^{2}\left( \nu _{\eta }^{2}-\nu _{\rho }^{2}\right) \right] }.
\label{mix}
\end{eqnarray}

The solution of the renormalization group at the lowest one-loop order gives
the running coupling constant for $\widetilde{M}\leq \mu $%
\begin{equation}
g_{i}^{-2}(\widetilde{M})=g_{i}^{-2}(\mu )+\frac{b_{i}}{8\pi ^{2}}\ln \left( 
\frac{\mu }{\widetilde{M}}\right) ,  \label{runn-coup-const}
\end{equation}

\noindent for $i=1,2,3,$ each one corresponding to the constant coupling of $%
U(1)_{Y},$ $SU(2)_{L}$ and $SU(3)_{c},$ respectively. Specifically, we use
the matching condition for the constant couplings, where the $SU(3)_{L}$
constant $g_{L}$ is the same as the $SU(2)_{L}$ constant$,$ i.e. $%
g_{2}=g_{L}.$ Running the constants at the scale $\mu =M_{Z^{\prime }},$ we
obtain for $g_{1}=g_{Y}$ and $g_{2}=g_{L}$

\begin{eqnarray}
g_{Y}^{2}(M_{Z^{\prime }}) &=&\frac{g_{Y}^{2}(\widetilde{M})}{1-\frac{b_{1}}{%
8\pi ^{2}}g_{Y}^{2}(\widetilde{M})\ln \left( \frac{M_{Z^{\prime }}}{%
\widetilde{M}}\right) }  \notag \\
g_{L}^{2}(M_{Z^{\prime }}) &=&\frac{g_{L}^{2}(\widetilde{M})}{1-\frac{b_{2}}{%
8\pi ^{2}}g_{L}^{2}(\widetilde{M})\ln \left( \frac{M_{Z^{\prime }}}{%
\widetilde{M}}\right) }  \label{g-running}
\end{eqnarray}

\noindent where the renormalization coefficients for a general $SU(N)$ gauge
group are defined by

\begin{equation}
b_{i}=\frac{2}{3}\sum_{fermions}Tr\left( T_{i}(F)T_{i}(F)\right) +\frac{1}{3}%
\sum_{scalars}Tr\left( T_{i}(S)T_{i}(S)\right) -\frac{11}{3}C_{2i}(G),
\label{renorm-coef1}
\end{equation}

\noindent with $T_{i}(I)$ corresponding to the generators of the fermionic ($%
F$) or scalar ($S$) representations. For $SU(N)$, $Tr(T(I)T(I))=1/2$ and $%
C_{2}(G)=N.$ For $U(1)_{Y},Tr(T(I)T(I))=Y^{2}$ and $C_{2}(G)=0.$ With the
above definitions, we can obtain the running Weinberg angle at the $%
Z^{\prime }$ scale

\begin{eqnarray}
S_{W}^{2}(M_{Z^{\prime }}) &=&\frac{g_{Y}^{2}(M_{Z^{\prime }})}{%
g_{L}^{2}(M_{Z^{\prime }})+g_{Y}^{2}(M_{Z^{\prime }})}  \notag \\
&=&S_{W}^{2}(\widetilde{M})\left[ \frac{1-\frac{b_{2}}{2\pi }\alpha _{2}(%
\widetilde{M})\ln \left( M_{Z^{\prime }}/\widetilde{M}\right) }{1-\frac{%
b_{1}+b_{2}}{2\pi }\alpha (\widetilde{M})\ln \left( M_{Z^{\prime }}/%
\widetilde{M}\right) }\right] .  \label{sw-running}
\end{eqnarray}

The running Weinberg angle depends on the particle content of the model.
First, we consider only the particle content of the effective Two Higgs
Doublet Model (THDM) below $M_{Z^{\prime }}$, where the heavy exotic
fermions $J_{m^{\ast }}$, $J_{3}$, $E_{j}$, the scalar singlets $\rho
_{3}^{++},\eta _{3}^{+}$ and the scalar triplet $\chi $ in Table \ref%
{tab:espectro} are not considered as active degrees of freedom i.e. they are
decoupled below the symmetry breaking scale $\mu _{331}$. Thus, the
renormalization coefficient takes the values

\begin{eqnarray}
b_{1} &=&\frac{20}{9}N_{f}+\frac{1}{6}N_{H}=7,  \notag \\
b_{2} &=&\frac{4}{3}N_{f}+\frac{1}{6}N_{H}-\frac{22}{3}=-3,
\label{renorm-coef}
\end{eqnarray}%
where $N_{f}=3$ is the number of fermion families and $N_{H}=2$ the number
of $SU(2)_{L}$ scalar doublets of the effective THDM. Taking $\widetilde{M}%
=M_{Z}$ we run the Eqs. \ref{g-running} and \ref{sw-running} using the
following values at the $M_{Z}$ scale

\begin{eqnarray}
\alpha _{Y}^{-1}(M_{Z}) &=&98.36461\pm 0.06657;\quad \alpha
^{-1}(M_{Z})=127.934\pm 0.027;  \notag \\
\alpha _{2}^{-1}(M_{Z}) &=&29.56938\pm 0.00068;\quad
S_{W}^{2}(M_{Z})=0.23113\pm 0.00015.  \label{Z-values}
\end{eqnarray}

Fig. 1 display the evolution of the Weinberg angle for the minmal 331
bilepton model. We can see a Landau pole at the energy scale $\mu \approx 4$
TeV, which corresponds to the lowest bound of the $Z^{\prime }$ mass for the
bilepton model, such as studied in ref. \cite{family-dependence}. Then, as
shown in ref. \cite{pole1}, perturbation theory can not be used at the TeV
scale energies in the minimal model even thought supersymmetry were
implemented.

\subsection{The extended 331 bilepton model (EM)}

In order to avoid the nonperturbative character of this model at the TeV
energies, it was proposed in ref. \cite{pole2} to introduce an additional
exotic particle content, where three lepton triplets with null hypercharge,
one scalar triplet with null hypercharge and two scalar doublets with $%
Y^{2}=9 $ remains as new degrees of freedom at energies below the scale $\mu
_{331}$ in order to obtain a perturbative regime at the $Z^{\prime }$ energy
scale. In this case, the renormalization coefficients take the form \cite%
{pole2}

\begin{eqnarray}
b_{1} &=&\frac{20}{9}N_{f}+\frac{1}{6}N_{H}+\frac{Y_{S}^{2}}{6}%
N_{S}+Y_{TF}^{2}N_{TF}+\frac{Y_{TS}^{2}}{4}N_{TS}=10,  \notag \\
b_{2} &=&\frac{4}{3}N_{f}+\frac{1}{6}N_{H}-\frac{22}{3}+\frac{8}{3}N_{TF}+%
\frac{2}{3}N_{TS}+\frac{1}{6}N_{S}=6,  \label{renorm-coef2}
\end{eqnarray}%
with $N_{S}=2$ the number of new scalar doublets, $N_{TF}=3$ the number of
new fermionic triplets, and $N_{TS}=1$ the number of new scalar triplets,
each one with the corresponding hypercharge values $Y_{S}=\pm 3,Y_{TF}=0$
and $Y_{TS}=0.$ In this case, the Weinberg angle runs as shown by Fig. 2 
\cite{pole2} exhibiting a decreasing behavior at the TeV scale. Thus the
perturbative character of the model is restored. The new particles are
embedded in three octet representations for leptons and one octet for the
scalar sector. Thus, there are additional new leptons that can be made heavy
through the scalar octet. The $U(1)_{X}$ quantum number of these new
multipets are null, so that they do not contribute to the anomalies and the
cancellation of anomalies is preserved. The Table \ref{tab:extended} resumes
the new bunch of particles of the extended 331 bilepton model.

\begin{table}[tbp]
\begin{center}
\begin{equation*}
\begin{tabular}{cc}
\hline
& $representation$ \\ \hline\hline
Leptons & $\ \Xi _{a}=\left( 
\begin{array}{ccc}
\frac{1}{\sqrt{2}}t_{a}^{0}+\frac{1}{\sqrt{6}}\lambda _{a}^{0} & t_{a}^{+} & 
\delta _{a}^{-} \\ 
t_{a}^{-} & \frac{-1}{\sqrt{2}}t_{a}^{0}+\frac{1}{\sqrt{6}}\lambda _{a}^{0}
& \delta _{a}^{--} \\ 
\xi _{a}^{+} & \xi _{a}^{++} & \frac{-2}{\sqrt{6}}\lambda _{a}^{0}%
\end{array}%
\right) \ $ \\ \hline\hline
Scalars & $\Sigma =\left( 
\begin{array}{ccc}
\frac{1}{\sqrt{2}}\phi ^{0}+\frac{1}{\sqrt{6}}\varphi ^{0} & \phi ^{+} & 
\varphi _{1}^{-} \\ 
\phi ^{-} & \frac{-1}{\sqrt{2}}\phi ^{0}+\frac{1}{\sqrt{6}}\varphi ^{0} & 
\varphi _{1}^{--} \\ 
\varphi _{2}^{+} & \varphi _{2}^{++} & \frac{-2}{\sqrt{6}}\varphi ^{0}%
\end{array}%
\right) $ \\ \hline
\end{tabular}%
\end{equation*}%
\end{center}
\caption{\textit{Additional particle content in the extended 331 bilepton
model (EM), with }$a=1,2,3$ in the lepton sector.\textit{\ }}
\label{tab:extended}
\end{table}

\subsection{Modification to the extended model (MEM)}

In the previous models, the usual exotic spectrum of the MM is considered as
heavy particles that are decoupled below the breaking scale\ $\mu _{331}$
and do not participate as active degree of freedom in the renormalization
coefficient. However, by comparing the data with radiative corrections to
the decay $Z\rightarrow b\overline{b}$ carried out in ref \cite%
{Roberto-Sampayo}, exotic quarks with a mass in the range $1.5-4$ TeV are
found for the 331-bilepton model, which lies in the range below the $%
Z^{\prime }$ scale. Then, we take the full 331 spectrum from tables \ref%
{tab:espectro} and \ref{tab:quince}, where the exotic particles are
considered as a degenerated spectrum. In agreement with the ref. \cite%
{Roberto-Sampayo}, it is reasonable to estimate $m_{J_{j},E_{j},\rho
_{3}^{++},\eta _{3}^{+},\chi _{1}^{-},\chi _{2}^{--}}\approx 3$ $TeV,$
defining a new running scale. Then, the Weinberg angle evolve from the $Z$
scale in a two stage process. First, below $\mu =3$ TeV, the running
parameters evolve as described by Eq. \ref{sw-running} with the coefficients
from Eq. \ref{renorm-coef} of the effective THDM, from where we obtain at 3
TeV that

\begin{eqnarray}
\alpha _{Y}^{-1}(3\text{ }TeV) &=&94.4724;\quad \alpha ^{-1}(3\text{ }%
TeV)=125.710;  \notag \\
\alpha _{2}^{-1}(3\text{ }TeV) &=&31.2374;\quad S_{W}^{2}(3\text{ }%
TeV)=0.24849.  \label{renorm-coef4}
\end{eqnarray}%
Later, above the 3 TeV threshold, the spectrum contains the exotic 331
particles with the following coefficients

\begin{eqnarray}
b_{1} &=&\frac{20}{9}N_{f}+\frac{1}{6}N_{H}+\frac{2}{3}\sum_{\text{sing.}%
}Y_{f}^{2}+\frac{1}{3}\sum_{\text{sing}}Y_{s}^{2}+\frac{2}{3}Y_{\chi }^{2}=%
\frac{79}{2},  \notag \\
b_{2} &=&\frac{4}{3}N_{f}+\frac{1}{6}N_{H}-\frac{43}{3}=-\frac{17}{6},
\label{renorm-coef3}
\end{eqnarray}%
with $Y_{f,s}$ the hypercharge of the singlets $f=$ $J_{j},E_{j}$ and $%
s=\rho _{3}^{++},\eta _{3}^{+}$, and $Y_{\chi }$ the hypercharge of the
doublet $\left( \chi _{1}^{-},\chi _{2}^{--}\right) .$ For the second stage,
we use the inputs at 3 TeV from Eq. \ref{renorm-coef4}. In this way, we
obtain the plot shown in Fig. 3. We can see how the behavior of the running
angle changes at the 3 TeV scale. The direct consequence to download the
exotic 331 spectrum at $3$ TeV $<$ $M_{Z^{\prime }}$ is that the Weinberg
angle increases faster and reachs to the Landau pole at 3.5 TeV below the $%
Z^{\prime }$ scale, which can be observe in the Fig. 3. In order to pull
this limit, we add into the degenerated exotic spectrum at 3 TeV scale of
the EM, the new particle content in Table \ref{tab:extended}. We obtain the
plot displayed in Fig. 4, where the Landau pole is pulled to 3.7 TeV, that,
however, does not restore the perturbative behavior near the $Z^{\prime }$
scale.

On the other hand, we can consider that the extended exotic particles from
Tab. 4 runs below the 3 TeV threshold. In this case, the evolution of the
coupling constant is controlled by the coefficients in Eq. \ref{renorm-coef2}%
, from where we get the new inputs at the threshold

\begin{eqnarray}
\alpha _{Y}^{-1}(3\text{ }TeV) &=&92.8048;\quad \alpha ^{-1}(3\text{ }%
TeV)=119.038;  \notag \\
\alpha _{2}^{-1}(3\text{ }TeV) &=&26.2334;\quad S_{W}^{2}(3\text{ }%
TeV)=0.22038.  \label{renorm-coef5}
\end{eqnarray}

Above 3 TeV, the full exotic spectrum of the MM is activated, and the
renormalization coefficients are defined as the combination of \ref%
{renorm-coef2} and \ref{renorm-coef3}, obtaining $b_{1}=85/2$ and $%
b_{2}=37/6.$ We see in this case that the Landau pole is pulled at very high
scales as we can observe in Fig. 5, where the Weinberg angle decreases with
the energy until the 3 TeV threshold, from where it begins to increase
slowly so that perturbation theory can be applied at the $Z^{\prime }$
scale. Thus, the Landau pole is pulled beyond the TeV energy scales.

\section{The $Z^{\prime }$ decays in the perturbative models \label{Z'-decay}%
}

Decay widths of $Z^{\prime }$ with and without flavor changing neutral
currents are obtained in ref. \cite{z2-decay} for models with $\beta =1/%
\sqrt{3}$, which displays a $Z^{\prime }$ mass bound of the order of $1.5$
TeV and does not exhibit Landau pole at the TeV energy scale. In this
section, we obtain the $Z^{\prime }$ decay for the bilepton model under the
framework of two different particle contents.

In Tab. \ref{tab:combination}, which is written in weak eigenstates $f^{0(r)}
$, we consider linear combinations in each representation $r=A,B,C$ among
the three families to obtain couplings in mass eigenstates $f^{(r)}$ through
the rotation matrix $R_{f}$ that diagonalize the Yukawa mass terms, where $%
f^{0(r)}=R_{f}f^{(r)}.$ Thus, the neutral couplings associated with $%
Z^{\prime }$ and the SM-fermions in mass eigenstates can be written as \cite%
{z2-decay}

\begin{eqnarray}
\mathcal{L}^{NC} &=&\frac{g_{L}}{2C_{W}}\left\{ \left[ \overline{U}\gamma
_{\mu }\left( \widetilde{\mathfrak{g}}_{v}^{U(r)}-\widetilde{\mathfrak{g}}%
_{a}^{U(r)}\gamma _{5}\right) U+\overline{D}\gamma _{\mu }\left( \widetilde{%
\mathfrak{g}}_{v}^{D(r)}-\widetilde{\mathfrak{g}}_{a}^{D(r)}\gamma
_{5}\right) D\right. \right.  \notag \\
&&+\left. \left. \overline{N}\gamma _{\mu }\left( \widetilde{g}_{v}^{N}-%
\widetilde{g}_{a}^{N}\gamma _{5}\right) N+\overline{E}\gamma _{\mu }\left( 
\widetilde{g}_{v}^{E}-\widetilde{g}_{a}^{E}\gamma _{5}\right) E\right]
Z^{\prime \mu }\right\} ,  \label{lag-5}
\end{eqnarray}%
with $U=\left( u,c,t\right) ^{T},$ $D=\left( d,s,b\right) ^{T},$ $N=\left(
\nu _{e},\nu _{\mu },\nu _{\tau }\right) ^{T},$ $E=\left( e,\mu ,\tau
\right) ^{T},$ and $\widetilde{\mathfrak{g}}_{v,a}^{q(r)}=R_{q}^{\dag }%
\widetilde{g}_{v,a}^{q(r)}R_{q}.$ Thus, we obtain flavor changing couplings
in the quark sector due to the matrix $R_{f}$. The vector and axial vector
couplings of the $Z^{\prime }$ field are listed in Table \ref{EW-couplings}
for each component.

\begin{table}[tbp]
\begin{center}
\begin{tabular}{|c|c|c|}
\hline
$Fermion$ & $\widetilde{g}_{v}^{f}$ & $\widetilde{g}_{a}^{f}$ \\ \hline\hline
$\nu _{j}$ & $\frac{-\sqrt{1-4S_{W}^{2}}}{2\sqrt{3}}$ & $\frac{-\sqrt{%
1-4S_{W}^{2}}}{2\sqrt{3}}$ \\ \hline
$e_{j}$ & $\frac{-1+10S_{W}^{2}}{2\sqrt{3}\sqrt{1-4S_{W}^{2}}}$ & $\frac{-%
\sqrt{1-4S_{W}^{2}}}{2\sqrt{3}}$ \\ \hline
$E_{j}$ & $\frac{1-5S_{W}^{2}}{\sqrt{3}\sqrt{1-4S_{W}^{2}}}$ & $\frac{%
1-S_{W}^{2}}{\sqrt{3}\sqrt{1-4S_{W}^{2}}}$ \\ \hline
$d_{m^{\ast }}$ & $\frac{1}{2\sqrt{3}\sqrt{1-4S_{W}^{2}}}$ & $\frac{\sqrt{%
1-4S_{W}^{2}}}{2\sqrt{3}}$ \\ \hline
$u_{m^{\ast }}$ & $\frac{1-6S_{W}^{2}}{2\sqrt{3}\sqrt{1-4S_{W}^{2}}}$ & $%
\frac{1+2S_{W}^{2}}{2\sqrt{3}\sqrt{1-4S_{W}^{2}}}$ \\ \hline
$J_{m^{\ast }}$ & $\frac{-1+9S_{W}^{2}}{\sqrt{3}\sqrt{1-4S_{W}^{2}}}$ & $%
\frac{-1+S_{W}^{2}}{\sqrt{3}\sqrt{1-4S_{W}^{2}}}$ \\ \hline
$u_{3}$ & $\frac{-1-4S_{W}^{2}}{2\sqrt{3}\sqrt{1-4S_{W}^{2}}}$ & $\frac{-%
\sqrt{1-4S_{W}^{2}}}{2\sqrt{3}}$ \\ \hline
$d_{3}$ & $\frac{-1+2S_{W}^{2}}{2\sqrt{3}\sqrt{1-6S_{W}^{2}}}$ & $\frac{%
-1-2S_{W}^{2}}{2\sqrt{3}\sqrt{1-4S_{W}^{2}}}$ \\ \hline
$J_{3}$ & $\frac{1-11S_{W}^{2}}{\sqrt{3}\sqrt{1-4S_{W}^{2}}}$ & $\frac{%
1-S_{W}^{2}}{\sqrt{3}\sqrt{1-4S_{W}^{2}}}$ \\ \hline
\end{tabular}%
\end{center}
\caption{\textit{Vector and axial vector couplings of fermions and }$%
Z^{\prime }$\textit{\ boson}.}
\label{EW-couplings}
\end{table}

As discussed in ref. \cite{z2-decay}, we can adopt the ansatz of Matsuda 
\cite{matsuda} on the texture of the quark mass matrices with the following
rotatation matrices

\begin{equation}
R_{D}=\left( 
\begin{array}{ccc}
c & s & 0 \\ 
-\frac{s}{\sqrt{2}} & \frac{c}{\sqrt{2}} & -\frac{1}{\sqrt{2}} \\ 
-\frac{s}{\sqrt{2}} & \frac{c}{\sqrt{2}} & \frac{1}{\sqrt{2}} \\ 
&  & 
\end{array}%
\right) ;\text{ }R_{U}=\left( 
\begin{array}{ccc}
c^{\prime } & 0 & s^{\prime } \\ 
-\frac{s^{\prime }}{\sqrt{2}} & -\frac{1}{\sqrt{2}} & \frac{c^{\prime }}{%
\sqrt{2}} \\ 
-\frac{s^{\prime }}{\sqrt{2}} & \frac{1}{\sqrt{2}} & \frac{c^{\prime }}{%
\sqrt{2}} \\ 
&  & 
\end{array}%
\right) ,  \label{rot-matrix}
\end{equation}%
with

\begin{eqnarray}
c &=&\sqrt{\frac{m_{s}}{m_{d}+m_{s}}};\qquad s=\sqrt{\frac{m_{d}}{m_{d}+m_{s}%
}};  \notag \\
c^{\prime } &=&\sqrt{\frac{m_{t}}{m_{t}+m_{u}}};\qquad s^{\prime }=\sqrt{%
\frac{m_{u}}{m_{t}+m_{u}}},  \label{rot-coef}
\end{eqnarray}%
where the quark masses at the $M_{Z^{\prime }}$ scale are

\begin{eqnarray}
\overline{m_{u}}(M_{Z^{\prime }}) &=&0.00170\text{ GeV};\qquad \overline{%
m_{c}}(M_{Z^{\prime }})=0.509\;\text{GeV},  \notag \\
\overline{m_{t}}(M_{Z^{\prime }}) &=&142.592\text{ GeV;\qquad }\overline{%
m_{d}}(M_{Z^{\prime }})=0.00342\;\text{GeV};  \notag \\
\overline{m_{s}}(M_{Z^{\prime }}) &=&0.0682\;\text{GeV};\qquad \overline{%
m_{b}}(M_{Z^{\prime }})=2.317\;\text{GeV}.  \label{running-quarks-mass}
\end{eqnarray}

\subsection{$Z^{\prime }$ decays in the extended model}

To calculate the decay widths, we must include the additional exotic
spectrum of the model from Sec. 2.2 in order to have a pertubative model at
the required energy level. The partial decay widths of $Z^{\prime }$ into
fermions $f\overline{f}$ at tree level are described by \cite{one, pitch}:

\begin{equation}
\Gamma _{Z^{\prime }\rightarrow \overline{f}f}^{0}=\frac{g_{L}^{2}M_{Z^{%
\prime }}N_{c}^{f}}{48\pi C_{W}^{2}}\sqrt{1-\mu _{f}^{\prime 2}}\left[
\left( 1+\frac{\mu _{f}^{\prime 2}}{2}\right) \left( \widetilde{g}%
_{v}^{f}\right) ^{2}+\left( 1-\mu _{f}^{\prime 2}\right) \left( \widetilde{g}%
_{a}^{f}\right) ^{2}\right] R_{QED}R_{QCD},  \label{Z'-tree}
\end{equation}

\noindent where $N_{c}^{f}=1$, 3 for leptons and quarks, respectively, $%
R_{QED,QCD}$ are global final-state QED and QCD corrections, and $\mu
_{f}^{\prime 2}=4m_{f}^{2}/M_{Z^{\prime }}^{2}$ takes into account
kinematical corrections only important for the top quark. In order to
calculate the corrections $R_{QED,QCD}$ defined in App. A, we use the
running QED ($\alpha $) and QCD ($\alpha _{s}$) constants at the $%
M_{Z^{\prime }}$ scale \cite{z2-decay,quarks-mass}. We can also generate $%
Z^{\prime }$ decays into different flavors of quarks through the Lagrangian in Eq.
\ref{lag-5}, where at tree level

\begin{equation}
\Gamma _{Z^{\prime }\rightarrow qq^{\prime }}^{0}=\frac{g_{L}^{2}M_{Z^{%
\prime }}}{16\pi C_{W}^{2}}\left[ \left( \widetilde{\mathfrak{g}}%
_{v}^{qq^{\prime }}\right) ^{2}+\left( \widetilde{\mathfrak{g}}%
_{a}^{qq^{\prime }}\right) ^{2}\right] R_{QED}R_{QCD}.  \label{fcnc}
\end{equation}

The decays into heavy exotic particles $Z^{\prime }\rightarrow
K^{+}K^{-},K^{++}K^{--},\eta _{3}^{+}\eta _{3}^{-},\rho _{3}^{++}\rho
_{3}^{--},J_{j},E_{j}$ are suppressed because in this case we consider that
only the effective THDM contributes, while the exotic particles are taken as
heavy particles that are decoupled at energies below the scale $\mu _{331}$.
Decays into light bosons $Z^{\prime }\rightarrow h^{0}h^{0},Z_{1}\gamma
,Z_{1}Z_{1},Z_{1}h^{0},W^{+}W^{-}$ which include the like-SM Higgs boson $%
h^{0}$ of the effective THDM are possible only through the small $%
Z-Z^{\prime }$ mixing angle $\theta $ defined by Eq. \ref{mix}, so that they
are very suppressed, and do not contribute to the branching ratios in a
significant amount \cite{tavares}. Then, we do not take into account these
decays in our analysis. On the other hand, since the oblique corrections are
virtual processes sensitive to the heavy particles at any scale of energy,\
we must consider the one loop corrections to the $Z^{\prime }$ decay due to
the exotic quarks $J_{j}$ and leptons $E_{j}$ at the scale $\mu
_{331}\gtrsim M_{Z^{\prime }}.$ A detailed study of these corrections is
perfomed in ref. \cite{z2-decay} in the framework of models with $\beta =1/%
\sqrt{3}$, where the $Z^{\prime }$ self-energy $\Sigma _{Z^{\prime
}Z^{\prime }},$ the $Z^{\prime }-Z$ vacuum polarization $\Pi _{Z^{\prime
}Z}, $ and the $Z^{\prime }-$photon vacuum polarization $\Pi _{Z^{\prime
}\gamma } $ (see the expressions in App. A) leads to the following decay width

\begin{equation}
\Gamma _{Z^{\prime }\rightarrow \overline{f}f}=\Gamma _{Z^{\prime
}\rightarrow \overline{f}f}^{0}\left( 1-\Delta _{f}^{\prime }\right) ,
\label{Z'-decay2}
\end{equation}%
with

\begin{equation}
\Delta _{f}^{\prime }\approx \frac{2\left( \widetilde{g}_{v}^{f}\Delta 
\widetilde{g}_{v}^{f}+\widetilde{g}_{a}^{f}\Delta \widetilde{g}%
_{a}^{f}\right) }{\left( \widetilde{g}_{v}^{f}\right) ^{2}+\left( \widetilde{%
g}_{a}^{f}\right) ^{2}},  \label{ew-corrections}
\end{equation}%
that contains the oblique radiative corrections in $\Delta \widetilde{g}%
_{v,a}^{f},$ which is defined as \cite{z2-decay}

\begin{eqnarray}
\Delta \widetilde{g}_{v}^{f} &\approx &2S_{W}C_{W}Q_{f}\Pi _{Z^{\prime
}\gamma }(M_{Z^{\prime }}^{2})+g_{v}^{f}\Pi _{Z_{2}Z_{1}}(M_{Z^{\prime
}}^{2})\left( 1+\frac{M_{Z}^{2}}{M_{Z^{\prime }}^{2}}\right) +\frac{1}{2}%
\widetilde{g}_{v}^{f}\Sigma _{Z^{\prime }Z^{\prime }}^{\prime
}(M_{z_{2}}^{2}),  \notag \\
\Delta \widetilde{g}_{a}^{f} &\approx &g_{a}^{f}\Pi _{Z^{\prime
}Z}(M_{Z^{\prime }}^{2})\left( 1+\frac{M_{Z}^{2}}{M_{Z^{\prime }}^{2}}%
\right) +\frac{1}{2}\widetilde{g}_{a}^{f}\Sigma _{Z^{\prime }Z^{\prime
}}^{\prime }(M_{Z^{\prime }}^{2}),  \label{effect-correction}
\end{eqnarray}%
where $\Sigma _{Z^{\prime }Z^{\prime }}^{\prime }=d\Sigma _{Z^{\prime
}Z^{\prime }}/dM_{Z^{\prime }}^{2}.$ The FCNC also exhibit corrections due
to the $Z^{\prime }$ self-energy, where

\begin{equation}
\Gamma _{Z^{\prime }\rightarrow qq^{\prime }}=\Gamma _{Z^{\prime
}\rightarrow qq^{\prime }}^{0}\left( 1-\Delta _{qq^{\prime }}^{\prime
}\right) ,  \label{fcnc-oneloop}
\end{equation}%
with

\begin{equation}
\Delta _{qq^{\prime }}^{\prime }\approx \frac{2\left( \widetilde{g}%
_{v}^{qq^{\prime }}\Delta \widetilde{g}_{v}^{qq^{\prime }}+\widetilde{g}%
_{a}^{qq^{\prime }}\Delta \widetilde{g}_{a}^{qq^{\prime }}\right) }{\left( 
\widetilde{g}_{v}^{qq^{\prime }}\right) ^{2}+\left( \widetilde{g}%
_{a}^{qq^{\prime }}\right) ^{2}},
\end{equation}%
and

\begin{equation}
\Delta \widetilde{g}_{v,a}^{qq^{\prime }}\approx \frac{1}{2}\widetilde{g}%
_{v,a}^{qq^{\prime }}\Sigma _{Z^{\prime }Z^{\prime }}^{\prime }(M_{Z^{\prime
}}^{2}).
\end{equation}

With the above considerations, and taking into account the running coupling
constants at the $Z^{\prime }$ resonance, we calculate the decay widths.
Using the Eqs. \ref{g-running} and \ref{sw-running} with the renormalization
coefficients from Eq. \ref{renorm-coef2} of the EM and the inputs from Eq. %
\ref{Z-values}, we get at the $M_{Z^{\prime }}\approx 4$ TeV scale that

\begin{eqnarray}
\alpha _{Y}^{-1}(M_{Z^{\prime }}) &=&92.3468;\text{\qquad }\alpha
^{-1}(M_{Z^{\prime }})=118.305;  \notag \\
\alpha _{2}^{-1}(M_{Z^{\prime }}) &=&25.9587;\qquad \alpha
_{s}^{-1}(M_{Z^{\prime }})=0.0779,  \notag \\
\overline{m_{t}}(M_{Z^{\prime }}) &=&142.592;\qquad S_{W}^{2}(M_{Z^{\prime
}})=0.2194.  \label{RCC}
\end{eqnarray}

With the above values and considering $M_{J_{j},E_{j}}=\mu _{331}\approx
M_{Z^{\prime }}=4$ TeV, we obtain the widths shown in table \ref%
{tab:run-quark-tree} from Eq. \ref{Z'-decay2}, and in Tab. \ref%
{tab:FCNC-Zprima-decay} for the FCNC contributions from Eq. \ref%
{fcnc-oneloop}. In the first case, values independent on the family
representation (universal of family) are obtained for the leptonic sector in
the two final rows in the table. In regard to the quark widths, we obtain
the family dependent decays shown, so that we identify three bilepton models
with the same particle content, but with observable differences. We also
obtain the branching ratios of each decay, where we assume that only decays
to SM particles are allowed, which is a reasonable approximation since other
two body decays are very suppressed. In the FCNC contributions, we obtain
very small values in the decay into the quarks $uc$ and $ut.$ However, we
obtain that the flavor changing decay into the quarks $ct$ and $sb$ for the $%
A$ and $B$ models, are about the same order of magnitude as the diagonal
components in Tab. \ref{tab:run-quark-tree}, so that they contribute to the
branching ratios. On the other hand, the model $C$ suppresses most of the
flavor changing effects of the neutral currents, which is also observed in
models with $\beta =1/\sqrt{3}$ \cite{z2-decay} (Studies on the FCNC in 331
models are also carried out in refs. \cite{tavares,Perez2,ghosal,chiang,sher}%
).

\begin{table}[tbp]
\begin{center}
\begin{tabular}{l|lll||lll}
\hline
\multicolumn{4}{c||}{$\Gamma _{ff}$ (GeV)} & \multicolumn{3}{c}{$Br$ ($%
Z^{\prime }\rightarrow ff$)$\times 10^{-2}$} \\ \hline\hline
& A & \multicolumn{1}{|l}{B} & \multicolumn{1}{|l||}{C} & A & 
\multicolumn{1}{|l}{B} & \multicolumn{1}{|l}{C} \\ \hline
$u\overline{u}$ & 73.862 & \multicolumn{1}{|l}{73.862} & 
\multicolumn{1}{|l||}{119.967} & 13.622 & \multicolumn{1}{|l}{13.635} & 
\multicolumn{1}{|l}{25.891} \\ \hline
$c\overline{c}$ & 73.862 & \multicolumn{1}{|l}{119.967} & 
\multicolumn{1}{|l||}{73.862} & 13.622 & \multicolumn{1}{|l}{22.146} & 
\multicolumn{1}{|l}{15.940} \\ \hline
$t\overline{t}$ & 119.962 & \multicolumn{1}{|l}{73.325} & 
\multicolumn{1}{|l||}{73.325} & 22.124 & \multicolumn{1}{|l}{13.536} & 
\multicolumn{1}{|l}{15.825} \\ \hline
$d\overline{d}$ & 34.211 & \multicolumn{1}{|l}{34.211} & 
\multicolumn{1}{|l||}{81.097} & 6.309 & \multicolumn{1}{|l}{6.315} & 
\multicolumn{1}{|l}{17.502} \\ \hline
$s\overline{s}$ & 34.211 & \multicolumn{1}{|l}{81.097} & 
\multicolumn{1}{|l||}{34.211} & 6.309 & \multicolumn{1}{|l}{14.971} & 
\multicolumn{1}{|l}{7.383} \\ \hline
$b\overline{b}$ & 81.097 & \multicolumn{1}{|l}{34.211} & 
\multicolumn{1}{|l||}{34.211} & 14.956 & \multicolumn{1}{|l}{6.315} & 
\multicolumn{1}{|l}{7.383} \\ \hline\hline
$\ell ^{+}\ell ^{-}$ &  & 38.732 &  & 7.143 & 7.150 & 8.359 \\ \hline
$\nu \overline{\nu }$ &  & 0.308 &  & 0.057 & 0.057 & 0.066 \\ \hline
\end{tabular}%
\end{center}
\caption{\textit{Partial widths and branching ratios of Z}$^{\prime }$ 
\textit{into fermions at one loop level for each representation A,B, and C,
in the framework of the EM. Leptons are universal of family.}}
\label{tab:run-quark-tree}
\end{table}

\begin{table}[tbp]
\begin{center}
\begin{tabular}{ccc}
\hline
$\Gamma _{qq^{\prime }}$ (GeV) & A-B & C \\ \hline\hline
$Z^{\prime }\rightarrow u\overline{c}$ & $5.00\times 10^{-4}$ & $0$ \\ \hline
$Z^{\prime }\rightarrow u\overline{t}$ & $5.00\times 10^{-4}$ & $2.00\times
10^{-3}$ \\ \hline
$Z^{\prime }\rightarrow c\overline{t}$ & $42.05$ & $0$ \\ \hline
$Z^{\prime }\rightarrow d\overline{s}$ & $1.91$ & $7.65$ \\ \hline
$Z^{\prime }\rightarrow d\overline{b}$ & $2.01$ & $0$ \\ \hline
$Z^{\prime }\rightarrow s\overline{b}$ & $40.02$ & $0$ \\ \hline
\end{tabular}%
\end{center}
\caption{\textit{Partial width of Z}$^{\prime }$ \textit{into quarks with
FCNC for each representation A,B, and C, in the framework of the EM.} }
\label{tab:FCNC-Zprima-decay}
\end{table}

\subsection{$Z^{\prime }$ decays in the modified extended model}

Now, we calculate the decay widths in the framework of the MEM from Sec.
2.3. However, we must take into account the new threshold defined by the
exotic quarks $J_{j}$ at 3 TeV, which modify the running QCD constant and
the mass of the top quark at the $Z^{\prime }$ scale. Thus, with the inputs
from Eq. \ref{renorm-coef5}, we obtain at $M_{Z^{\prime }}\approx 4$ TeV 
\begin{eqnarray}
\alpha _{Y}^{-1}(M_{Z^{\prime }}) &=&67.229;\text{\qquad }\alpha
^{-1}(M_{Z^{\prime }})=101.377;  \notag \\
\alpha _{2}^{-1}(M_{Z^{\prime }}) &=&22.5224;\qquad \alpha _{s}(M_{Z^{\prime
}})=0.0791,  \notag \\
\overline{m_{t}}(M_{Z^{\prime }}) &=&142.579;\qquad S_{W}^{2}(M_{Z^{\prime
}})=0.2222  \label{RCC2}
\end{eqnarray}

Compairing with Eq. \ref{RCC}, we note the strong dependence of the running
constants with the particle content. The running masses of the other quarks do
not present an appreciable change respect the values given by Eq. \ref%
{running-quarks-mass}. Taking into account the one loop corrections with $%
M_{J_{j},E_{j}}=3$ TeV, we obtain the widths from Eq. \ref{Z'-decay2} and
the branching ratios shown in Tab. \ref{tab:run-quark-tree2}. Although the
decays $Z_{2}\rightarrow K^{+}K^{-},K^{++}K^{--},\eta _{3}^{+}\eta
_{3}^{-},\rho _{3}^{++}\rho _{3}^{--},J_{j},E_{j}$ are not forbidden if we
download all the exotic spectrum at the TeV scales, they are very suppressed
by kinematical factors \cite{tavares}. Table \ref{tab:FCNC decays} show the
FCNC widths for this case.

\begin{table}[tbp]
\begin{center}
\begin{tabular}{l|lll||lll}
\hline
\multicolumn{4}{c||}{$\Gamma _{ff}$ (GeV)} & \multicolumn{3}{c}{$Br$ ($%
Z^{\prime }\rightarrow ff$)$\times 10^{-2}$} \\ \hline\hline
& A & \multicolumn{1}{|l}{B} & \multicolumn{1}{|l||}{C} & A & 
\multicolumn{1}{|l}{B} & \multicolumn{1}{|l}{C} \\ \hline
$u\overline{u}$ & 78.770 & \multicolumn{1}{|l}{78.770} & 
\multicolumn{1}{|l||}{128.91} & 13.660 & \multicolumn{1}{|l}{13.672} & 
\multicolumn{1}{|l}{25.974} \\ \hline
$c\overline{c}$ & 78.770 & \multicolumn{1}{|l}{128.91} & 
\multicolumn{1}{|l||}{78.770} & 13.660 & \multicolumn{1}{|l}{22.375} & 
\multicolumn{1}{|l}{15.872} \\ \hline
$t\overline{t}$ & 128.90 & \multicolumn{1}{|l}{78.200} & 
\multicolumn{1}{|l||}{78.200} & 22.353 & \multicolumn{1}{|l}{13.573} & 
\multicolumn{1}{|l}{15.757} \\ \hline
$d\overline{d}$ & 36.597 & \multicolumn{1}{|l}{36.597} & 
\multicolumn{1}{|l||}{85.817} & 6.346 & \multicolumn{1}{|l}{6.352} & 
\multicolumn{1}{|l}{17.292} \\ \hline
$s\overline{s}$ & 36.597 & \multicolumn{1}{|l}{85.817} & 
\multicolumn{1}{|l||}{36.597} & 6.346 & \multicolumn{1}{|l}{14.895} & 
\multicolumn{1}{|l}{7.374} \\ \hline
$b\overline{b}$ & 85.817 & \multicolumn{1}{|l}{36.597} & 
\multicolumn{1}{|l||}{36.597} & 14.882 & \multicolumn{1}{|l}{6.352} & 
\multicolumn{1}{|l}{7.374} \\ \hline\hline
$\ell ^{+}\ell ^{-}$ &  & 41.733 &  & 7.237 & 7.243 & 8.409 \\ \hline
$\nu \overline{\nu }$ &  & 0.310 &  & 0.054 & 0.054 & 0.062 \\ \hline
\end{tabular}%
\end{center}
\caption{\textit{Partial widths and branching ratios of Z}$^{\prime }$ 
\textit{into fermions at one loop level for each representation A,B, and C,
in the framework of the MEM. Leptons are universal of family.}}
\label{tab:run-quark-tree2}
\end{table}

\begin{table}[tbp]
\begin{center}
\begin{tabular}{ccc}
\hline
$\Gamma _{qq^{\prime }}$ (GeV) & A-B & C \\ \hline\hline
$Z^{\prime }\rightarrow u\overline{c}$ & $5.20\times 10^{-4}$ & $0$ \\ \hline
$Z^{\prime }\rightarrow u\overline{t}$ & $5.20\times 10^{-4}$ & $2.53\times
10^{-3}$ \\ \hline
$Z^{\prime }\rightarrow c\overline{t}$ & $43.63$ & $0$ \\ \hline
$Z^{\prime }\rightarrow d\overline{s}$ & $1.98$ & $9.36$ \\ \hline
$Z^{\prime }\rightarrow d\overline{b}$ & $2.08$ & $0$ \\ \hline
$Z^{\prime }\rightarrow s\overline{b}$ & $41.51$ & $0$ \\ \hline
\end{tabular}%
\end{center}
\caption{\textit{Partial width of Z}$^{\prime }$ \textit{into quarks with
FCNC for each representation A,B, and C for the MEM}}
\label{tab:FCNC decays}
\end{table}

\subsection{Discussion}

The results of the decay widths and branching ratios show observable
differences between the representation $A,B$ and $C,$ thus each case from
Table \ref{tab:combination} define three models that leads to different
predictions of the extra neutral boson $Z^{\prime }$. Then, the assignation
of the phenomenological quarks into the 331 representations define 3
different bilepton models with the same particle content. \ 

Comparing the above results with other estimations performed in the
framework of models with $\beta =1/\sqrt{3}$ \cite{z2-decay} and bilepton
models \cite{tavares}, we get partial widths about one order of magnitude
bigger for the quark sector and the charged lepton sector, while the
neutrinos results one order of magnitude smaller. These diferences in the
partial widths may be used in order to distinguish which 331 model could
describe in a better way the physics at the TeV energy scale in the future
experiments$.$ The branching ratios are in the range $10^{-2}\leq
Br(Z_{2}\rightarrow qq)\leq 10^{-1}$ for quarks, $Br(Z_{2}\rightarrow \ell
^{+}\ell ^{-})\approx 10^{-2}$ for the charged leptons, and $%
Br(Z_{2}\rightarrow \nu \nu )\approx 10^{-4}$ for neutrinos. Comparing with
the model $\beta =1/\sqrt{3}$ \cite{z2-decay} or the bilepton model from
ref. \cite{tavares}, we get branching ratios between one and two order of
magnitude smaller.

On the other hand, by comparing Tables \ref{tab:run-quark-tree} and \ref%
{tab:run-quark-tree2}, we get that the modified extended particle content
predict widths about 6 \% bigger that extended bilepton model. In general,
we see that each decay width increase when exotic particles, in particular
when the exotic quarks of the minimal bilepton model are downloaded below
the $Z^{\prime }$ resonance. The branching ratios present very similar
values between both particle contents, so that the additional effects of the
new threshold an the presence of additional exotic particles are canceled in
the calculations of the ratios. Just as the top quark contribute to the one
loop $Z$ decay, the above decay widths were calculated at one loop level
throught the heavy particles $J_{j}$ and $E_{j}$, where we take into account
the running coupling constant at the $M_{Z^{\prime }}$ energy level.

The branching ratios receive important contributios to the flavor changing
decays $Z^{\prime }\rightarrow ct,sb$ in the $A$ and $B$ models. The other
flavor changing effects are very small, in particular, the FCNC
phenomenology is practically absent for $C$ models, as can be seen in Tabs. %
\ref{tab:FCNC-Zprima-decay} and \ref{tab:FCNC decays}.

\ 

\section{Conclusions}

In this work we studied the main contributions of the $Z^{\prime }$ decay in
the framework of the 331 bilepton model with extra particles that pull the
Landau pole beyond the TeV scale energies in order to get a perturbative
model at the $Z^{\prime }$ peak. These calculations were performed for two
different particle contents, first in the model proposed in ref. \cite{pole2}
where three lepton triplets with null hypercharge, one scalar triplet with
null hypercharge and two scalar doublets with $Y^{2}=9$ remains as new
degrees of freedom at energies below $M_{Z^{\prime }},$ and later in a
modified extended model (MEM) where the full 331 spectrum from tables \ref%
{tab:espectro} and \ref{tab:quince} is taken into account with a degenerated
exotic spectrum at 3 $TeV$ in agreement with some phenomenological studies 
\cite{Roberto-Sampayo}$.$ In particular, the presence of the exotic quarks $%
J_{j}$ as an active degree of freedom below the $Z^{\prime }$ scale, define
a new threshold that change the evolution of the running parameters. In
general, we obtained predictions for 6 different perturbative models with
two particle contents. As seen in tables \ref{tab:run-quark-tree}-\ref%
{tab:FCNC decays}, each case $A,B$ and $C$ from Tab. \ref{tab:combination}
lead to different predictions of the $Z^{\prime }$ decays, thus they
represent three different bilepton model. The width decays obtained are of
the order of 10$^{2}$ GeV for quarks and charged leptons which are one order
of magnitude bigger than the predicted by the models with $\beta =1/\sqrt{3}%
, $ while the decays into neutrinos are of the order 10$^{-1},$ one order of
magnitude smaller than in models $\beta =1/\sqrt{3}.$ We also obtained the
branching ratios, where decays into exotic particles are not taken into
account by kinematical reasons and decays into light bosons are suppressed
by the small $Z-Z^{\prime }$ mixing angle. However, FCNC effects through the
ansatz of Matsuda on the texture of the quark mass matrices leads to
important contribution in the decays $Z^{\prime }\rightarrow ct,sb,$ which
are of the order of 10$^{2}$ GeV and should be taken into account in the
total decay width.

We acknowledge financial support from Colciencias. We also thanks to HELEN
(High Energy Physics LatinAmerican - European Network) Program

\section*{Appendix}

\appendix

\section{Radiative Corrections\label{appendixAA}}

\bigskip The $Z^{\prime }$ decay in Eq. (\ref{Z'-tree}) contains global QED
and QCD corrections through the definition of $R_{QED}=1+\delta _{QED}^{f}$
and $R_{QCD}=1+(1/2)\left( N_{c}^{f}-1\right) \delta _{QCD}^{f}$, where \cite%
{one, pitch}

\begin{eqnarray}
\delta _{QED}^{f} &=&\frac{3\alpha Q_{f}^{2}}{4\pi };  \notag \\
\delta _{QCD}^{f} &=&\frac{\alpha _{s}}{\pi }+1.405\left( \frac{\alpha _{s}}{%
\pi }\right) ^{2}-12.8\left( \frac{\alpha _{s}}{\pi }\right) ^{3}-\frac{%
\alpha \alpha _{s}Q_{f}^{2}}{4\pi ^{2}}  \label{QCD}
\end{eqnarray}%
with $\alpha $ and $\alpha _{s}$ the electromagnetic and QCD constants,
respectively. The values $\alpha $ and $\alpha _{s}$ are calculated at the $%
M_{Z^{\prime }}$ scale as shown in Eqs. \ref{RCC} and \ref{RCC2} in
agreement with each particle content and quark thresholds \cite{quarks-mass}%
.\ 

The correction due to the $Z^{\prime }$ self-energy leads to the
wavefunction renormalization%
\begin{equation}
Z^{\prime }\rightarrow Z_{R}^{\prime }\approx \left( 1-\frac{1}{2}\Sigma
_{Z^{\prime }Z^{\prime }}^{(fin)\prime }(q^{2})\right) Z^{\prime },
\label{Z'-renormalizado}
\end{equation}

\noindent which is sensitive to the heavy fermions $f_{j}=J_{j},E_{j},$ where

\begin{eqnarray}
\Sigma _{Z^{\prime }Z^{\prime }}^{(fin)^{\prime }}(q^{2}) &=&\frac{d\Sigma
_{Z^{\prime }Z^{\prime }}^{(fin)}}{dq^{2}}=\frac{1}{12\pi ^{2}}\left( \frac{%
g_{L}}{2C_{W}}\right) ^{2}\sum_{f_{j}}\left\{ \left( \widetilde{g}%
_{v}^{f_{j}}\right) ^{2}\left( \frac{2}{3}-\ln \frac{m_{f_{j}}^{2}}{q^{2}}%
\right) \right.  \notag \\
&&+\left. \left( \widetilde{g}_{a}^{f_{j}}\right) ^{2}\left( \frac{2}{3}-\ln 
\frac{m_{f_{j}}^{2}}{q^{2}}-\frac{6m_{f_{j}}^{2}}{q^{2}}\right) \right\} .
\label{derivative-Z'-Z'-self}
\end{eqnarray}

The \noindent $Z^{\prime }-Z$ self-energy leads to the following vacuum
polarization

\begin{eqnarray}
\Pi _{Z^{\prime }Z}^{(fin)}(q^{2}) &\approx &\frac{1}{12\pi ^{2}}\left( 
\frac{g_{L}}{2C_{W}}\right) ^{2}\sum_{f_{j}}\left\{ \widetilde{g}%
_{v}^{f_{j}}g_{v}^{f_{j}}\left[ -\ln \frac{m_{f_{j}}^{2}}{q^{2}}-\frac{1}{3}%
\right] \right.  \notag \\
&&+\left. \widetilde{g}_{a}^{f_{j}}g_{a}^{f_{j}}\left[ \left( \frac{%
6m_{f_{j}}^{2}}{q^{2}}-1\right) \ln \frac{m_{f_{j}}^{2}}{q^{2}}-\frac{1}{3}%
\right] \right\} ,  \label{Z'-Z-self}
\end{eqnarray}

\noindent and the $Z^{\prime }$-photon vacuum polarization is given by

\begin{equation}
\Pi _{Z^{\prime }\gamma }^{(fin)}(q^{2})\approx \frac{1}{12\pi ^{2}}\frac{%
g_{L}^{2}S_{W}}{2C_{W}}\sum_{f_{j}}\left\{ Q_{f_{j}}\widetilde{g}_{v}^{f_{j}}%
\left[ -\ln \frac{m_{f_{j}}^{2}}{q^{2}}-\frac{1}{3}\right] \right\} ,
\label{Z'-photon-self}
\end{equation}

\noindent with $Q_{f_{j}}$ the electric charge of the virtual $f_{j}$
fermions given in table \ref{tab:espectro}.

\newpage

\begin{figure}[tbph]
\centering \includegraphics[scale=1]{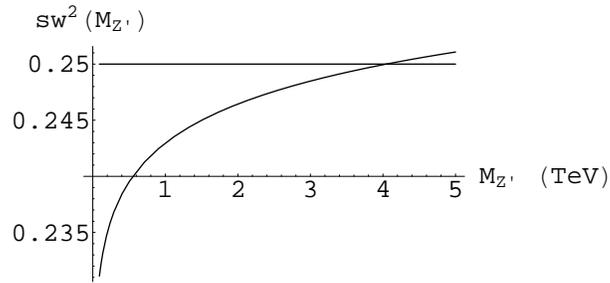}
\caption{Running Weinberg angle for the minimal 331 bilepton model (MM) in
the framework of the effective THDM (without exotic particles). A
Landau-like pole is found at the $M_{Z^{\prime}}\approx4$ TeV energy scale}
\label{figura1}
\end{figure}

\begin{figure}[tbph]
\centering \includegraphics[scale=1]{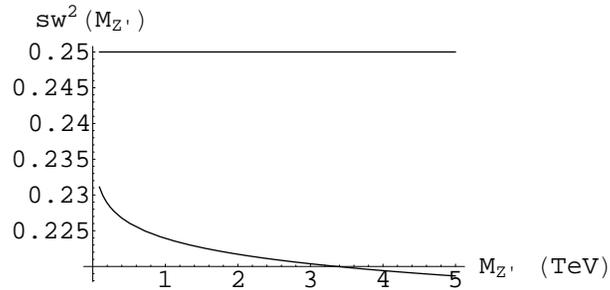}
\caption{Running Weinberg angle for the extended 331 bilepton model (EM)
with a new set of leptonic particles. The angle show decreasing behavior,
from where the Landau pole is avoid}
\label{figura2}
\end{figure}

\begin{figure}[tbph]
\centering \includegraphics[scale=1]{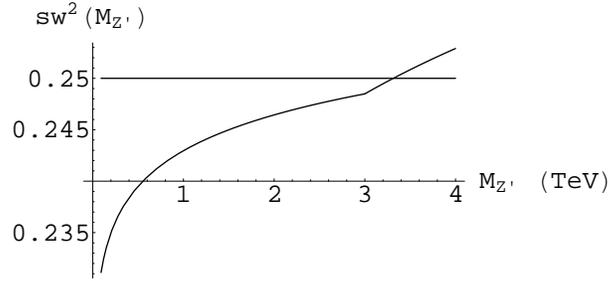}
\caption{Running Weinberg angle for the MM with exotic particles at the $3$
TeV threshold below $M_{Z^{\prime}}$ . A Landau-like pole is found at 3.5 TeV
energy scale}
\label{figura3}
\end{figure}

\begin{figure}[tbph]
\centering \includegraphics[scale=1]{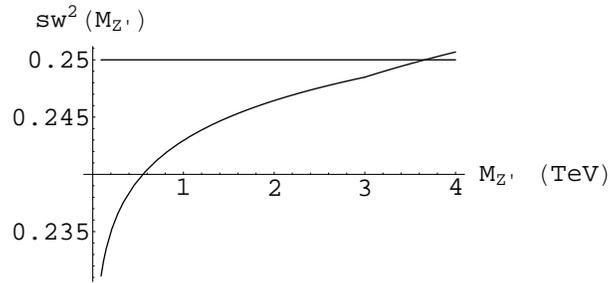}
\caption{Running Weinberg angle for the modified extended model (MEM) with
the full exotic spectrum of the MM and EM at the $3$ TeV threshold. A
Landau-like pole is found at the $M_{Z^{\prime}}\approx3.7$ TeV energy scale}
\label{figura4}
\end{figure}

\begin{figure}[tbph]
\centering \includegraphics[scale=1]{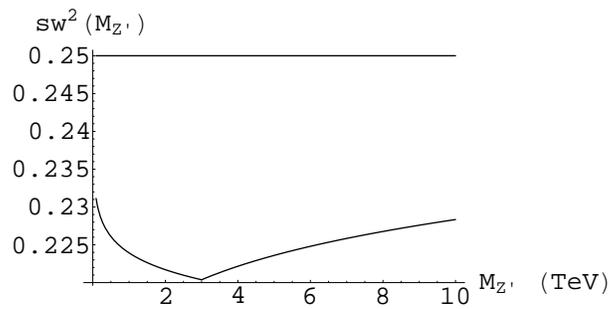}
\caption{Running Weinberg angle for the MEM with the new leptonic particles
running below the $3$ TeV threshold  defined by the minimal exotic spectrum.
A Landau-like pole is pulled far away from the TeV energy scales.}
\label{figura5}
\end{figure}

\end{document}